\documentclass[aps,prd,showpacs,preprintnumbers,superscriptaddress,nofootinbib]{revtex4}
\usepackage[dvips]{graphicx}
\usepackage{bm,latexsym,amsmath,amssymb,amsfonts}
\newcommand{\stac}[2]{\stackrel{\scriptscriptstyle {#1}}{#2}}
\newcommand*{\py}{\partial_y}
\newcommand*{\cD}{{\cal D}}
\newcommand*{\cF}{{\cal F}}
\newcommand*{\cJ}{{\cal J}}
\newcommand*{\cK}{{\cal K}}
\newcommand*{\e}{{\rm e}}
\begin{document}

\title{Low energy effective theory on a regularized brane in 6D gauged chiral supergravity}

\author{Frederico~Arroja}
\email[Email: ]{Frederico.Arroja"at"port.ac.uk}
\affiliation{Institute of Cosmology and Gravitation, University of Portsmouth, Portsmouth
PO1 2EG, UK}
\author{Tsutomu~Kobayashi}
\email[Email: ]{tsutomu"at"gravity.phys.waseda.ac.jp}
\affiliation{Department of Physics, Waseda University, Okubo 3-4-1, Shinjuku, Tokyo 169-8555, Japan}
\author{Kazuya~Koyama}
\email[Email: ]{Kazuya.Koyama"at"port.ac.uk}
\affiliation{Institute of Cosmology and Gravitation, University of Portsmouth, Portsmouth
PO1 2EG, UK}
\author{Tetsuya~Shiromizu}
\email[Email: ]{shiromizu"at"phys.titech.ac.jp}
\affiliation{Department of Physics, Tokyo Institute of Technology, Tokyo 152-8551, Japan}

\begin{abstract}
We derive the low energy effective theory on a brane in
six-dimensional chiral supergravity. The conical 3-brane
singularities are resolved by introducing cylindrical codimension
one 4-branes whose interiors are capped by a regular spacetime.
The effective theory is described by the Brans-Dicke (BD) theory
with the BD parameter given by $\omega_{\rm BD}=1/2$. The BD field
is originated from a modulus which is associated with the scaling
symmetry of the system. If the dilaton potentials on the branes
preserve the scaling symmetry, the scalar field has an exponential
potential in the Einstein frame. We show that the time dependent
solutions driven by the modulus in the four-dimensional effective
theory can be lifted up to the six-dimensional exact solutions
found in the literature. Based on the effective theory, we discuss
a possible way to stabilize the modulus to recover standard
cosmology and also study the implication for the cosmological
constant problem.

\end{abstract}

 \preprint{WU-AP/274/07} \maketitle

\section{Introduction}
Recently, much attention has been paid to six-dimensional
supergravity \cite{sugra, sugra1, sugra2}. The most intriguing
property of six-dimensional supergravity is that the
four-dimensional spacetime is always Minskowski even in the
presence of branes with tension. A 3-brane with tension induces
only a deficit angle in the six-dimensional spacetime and the
tension does not curve the four-dimensional spacetime within the
brane. This feature is called self-tuning and it may solve the
cosmological constant problem \cite{cc1, cc-rev}. This is the
basis of the supersymmetric large extra-dimension (SLED) proposal
\cite{burgess}.

There have been several objections to the idea of self-tuning
\cite{cc2, GP}. The self-tuning relies on the classical scaling
property of the model. The six-dimensional equations of motion are
invariant under the constant rescaling $g_{MN} \to e^{\omega}
g_{MN}$ and $e^{\phi} \to e^{\phi -\omega}$, where $g_{MN}$
denotes the six-dimensional metric and $\phi$ is the dilaton field. Then there
is a modulus associated with this scaling property. Ref.~\cite{GP}
derived an effective potential for this modulus. This modulus is
shown to have an exponential potential. Then there must be a
fine-tuning of parameters to ensure that the potential vanishes in
order to have a static solution. This is the reason why the static
solution always has vanishing cosmological constant. However, if
this fine-tuning is broken, the modulus acquires a runaway
potential and the four-dimensional spacetime becomes non-static.
Non-static solutions in six-dimensional supergravity have been
derived and they are supposed to correspond to the response of the
bulk geometry to a change of tension of branes
 \cite{Cline:2003ak, TBRH, KM2, CS}.

However, it is difficult to deal with an arbitrary change of
tension with a brane described by a pure conical singularity. This
is because if we put matter on the brane other than cosmological
constant, the metric diverges at the position of the brane.
Recently, it was suggested that we can regularize the brane by
resolving it by a codimension one cylindrical 4-brane
\cite{peloso1, ppz, peloso2, KM}. This type of models may be
regarded as a variation of Kaluza-Klein/hybrid brane world
\cite{hybrid}. Once the brane becomes a codimension one object, it
is possible to put arbitrary matter on the brane without having
the divergence of the metric. Then it becomes possible to study
the effect of the change of tension on the four-dimensional
geometry on the brane.

There is another interesting issue of whether it is possible to
recover conventional cosmology at low energies in six-dimensional
models. Recent works have shown that it is impossible to recover
sensible cosmology if one derives cosmological solutions by
considering a motion of branes in a given static bulk spacetime
\cite{ppz_cos, ml_cos}. It was concluded that the time-dependence
of the bulk spacetime should be taken into account.

In this paper, we derive a four-dimensional effective theory for
the modulus in six-dimensional supergravity with resolved 4-branes
by extending the analysis of Ref.~\cite{FKS} which studied the low
energy effective theory in the Einstein-Maxwell theory \cite{E-M}.
Arbitrary matter and potentials for the dilaton on 4-branes are
allowed to exist. We use the gradient expansion technique to solve
the six-dimensional geometry assuming that the deviation from the
static solution is small \cite{soda, sk}. The gradient expansion
method has been applied to various types of braneworlds \cite{ge}.
Using this method, it is possible to solve the non-trivial
dependence of the bulk geometry on the four-dimensional
coordinates. By solving the effective four-dimensional equations,
we can derive the time-dependent solutions and compare them with
the exact six-dimensional time dependent solutions found in the
literature \cite{TBRH, KM2, CS}. It is also possible to study
whether we can reproduce sensible cosmology at low energies or
not. We also study the possibility to stabilize the modulus using
the potentials for the dilaton on the branes along the line of
Ref.~\cite{uvcap}.

The paper is organized as follows. In section II, basic equations
are summarized. In section III, we solve the six-dimensional
equations of motion using the gradient expansion method. In
section IV, the effective theory on the regularized branes is
derived by imposing junction conditions. Then we derive time
dependent cosmological solutions in the effective theory and
compare them with the exact six-dimensional solutions. The
possible way to stabilize the modulus is discussed. Section V is
devoted to conclusions.

\section{Basic equations}\label{sec:BASIC}

The relevant part of the supergravity action we consider is
\begin{eqnarray}
S =\int d^6x\sqrt{-g}\left[\frac{M^4}{2}R
-\frac{M^4}{2}\left(\partial \phi\right)^2
-\frac{1}{4}F^2e^{-\phi}-\frac{M^4}{2L_I^2} e^\phi \right],
\end{eqnarray}
where $\phi$ is the dilaton, $M$ is the fundamental scale of
gravity, $(\partial\phi)^2:=g^{MN}\partial_M\phi\partial_N\phi$,
$F^2:=F_{MN}F^{MN}$, and $F_{MN}=\partial_MA_N-\partial_NA_M$ is
the field strength of the gauge field $A_M$. For the moment we are
interested in solving the 6D bulk equations of motion. In Sec.
\ref{sec:branes} we will add two 4-branes (at positions $y=y_\pm$)
and $L_I$ denotes the different bulk curvature scales on either
sides of the branes, see Fig. \ref{fig:fig}. We start with the
axisymmetric metric ansatz
\begin{eqnarray}
g_{MN}dx^Mdx^N
=L_I^2 e^{2\lambda(x)}\frac{dy^2}{f(y)}+\ell^2e^{2[\psi(y,x)-\lambda(x)]}f(y)d \theta^2+
2\ell b_{\mu}(y, x)d\theta dx^{\mu}+
a^2(y) \bar{h}_{\mu\nu}(y, x)dx^\mu dx^\nu,
\label{metric}
\end{eqnarray}
where capital Latin indices numerate the 6D coordinates while the
Greek indices are restricted to the 4D coordinates.

The evolution equations along the $y$-direction are given by
\begin{eqnarray}
n^y \py K_{\hat\mu}^{\;\hat\nu}+\hat K K_{\hat\mu}^{\;\hat\nu}
={}^{5}\!R_{\hat\mu}^{\;\hat\nu}-e^{-\lambda(x)}{}^5\!D_{\hat \mu}{}^5 \!D^{\hat \nu}e^{\lambda (x)}
-\partial_{\hat \mu} \phi\partial^{\hat \nu} \phi
-\frac{1}{4L_I^2} e^{\phi} \delta_{\hat\mu}^{\;\hat\nu}
-\frac{1}{M^4}\left(F_{\hat\mu M}F^{\hat\nu M}-\frac{1}{8}\delta_{\hat\mu}^{\;\hat\nu}F^2\right)e^{-\phi},
\end{eqnarray}
where $n^y=e^{-\lambda}{\sqrt {f}}/L_I$, $K_{\hat\mu}^{\;\hat\nu}$ is the extrinsic curvature
of $y=$ constant hypersurfaces, $\hat K$ is its 5D trace,
${}^{5}\!R_{\hat\mu}^{\;\hat\nu}$ is the 5D Ricci tensor
and ${}^{5}\! D_{\hat\mu}$ is the covariant derivative with respect to
the 5D metric. Here, $\hat\mu = \mu$ and $\theta$.
The Hamiltonian constraint is
\begin{eqnarray}
{}^{5}\!R+K_{\hat\mu}^{\;\hat\nu}K_{\hat\nu}^{\;\hat\mu}-\hat K^2
=-\frac{2}{M^4}\left(F_{yM}F^{yM}-\frac{1}{4}F^2\right)e^{-\phi}
-2(n^y \partial_y \phi)^2+(\partial \phi)^2+\frac{1}{L_I^2} e^\phi,
\end{eqnarray}
and the momentum constraints are
\begin{eqnarray}
{}^{5}\! D_{\hat\nu}\left(K_{\hat\mu}^{\;\hat\nu}-\delta_{\hat\mu}^{\;\hat\nu}\hat K\right)
=\frac{1}{M^4}F_{\hat\mu M}F^{yM}n_ye^{-\phi} +D_{\hat \mu} \phi \;n^y \partial_y \phi ,
\end{eqnarray}
where $n_y=e^{\lambda} L_I/{\sqrt {f}}$.

The Maxwell equations are given by
\begin{eqnarray}
\nabla_M\left(e^{-\phi}F^{MN}\right) =0,
\end{eqnarray}
where $\nabla_M$ is the covariant derivative with respect to the 6D metric.
The dilaton equation of motion is
\begin{eqnarray}
\nabla_M\nabla^M \phi+\frac{1}{4M^4} F^2e^{-\phi}-\frac{1}{2L_I^2}e^\phi=0.
\end{eqnarray}

\section{Gradient expansion approach}\label{sec:GE}

In this section we will use the gradient expansion method
\cite{soda, sk} to solve the 6D bulk equations. We assume that the
length scale $\ell$ is of the same order of $L_I$.  The small
expansion parameter is the ratio of the bulk curvature scale to
the 4D intrinsic curvature scale,
\begin{eqnarray*}
\varepsilon = \ell^2|R|.
\end{eqnarray*}
We expand the various quantities as
\begin{eqnarray}
\bar{h}_{\mu\nu}=h_{\mu\nu}(x)+\varepsilon h^{(1)}_{\mu\nu}(y,
x)+\cdots, \quad \psi = \psi^{(0)}+\varepsilon\psi^{(1)}+\cdots,
\quad \phi = \phi^{(0)}+\varepsilon\phi^{(1)}+\cdots,
\nonumber\\
K_{\mu}^{\;\nu}=\stac{(0)}{K_{\mu}^{\;\nu}}+\varepsilon
\!\stac{(1)}{K_{\mu}^{\;\nu}}+\cdots, \quad
K_{\theta}^{\;\theta}=\stac{(0)}{K_{\theta}^{\;\theta}}+\varepsilon
\!\stac{(1)}{K_{\theta}^{\;\theta}}+\cdots, \quad
F_{y\theta}=\stac{(0)}{F_{y\theta}}+\varepsilon\!\stac{(1)}{F_{y\theta}}+\cdots.
\end{eqnarray}
As to the other quantities, we follow~\cite{FKS} and first assume
\begin{eqnarray}
b_{\mu}=\varepsilon^{1/2}b^{(1/2)}_{\mu}+\cdots,
\quad
K_{\theta}^{\;\nu}=\varepsilon^{1/2}\!\stac{(1/2)}{K_{\theta}^{\;\nu}}+\cdots,
\nonumber\\
F^{\mu y}=\varepsilon^{1/2}\stac{(1/2)}{F^{\mu y}}+\cdots,
\quad
F_{\mu\nu}=\varepsilon\!\stac{(1)}{F_{\mu\nu}}+\cdots,
\end{eqnarray}
and then will show that all the ${\cal O}(\varepsilon^{1/2})$
quantities in fact vanish. Since $\partial_{\mu}A_{\theta}\sim
\varepsilon^{1/2}\py A_{\theta}$, we have
$F_{\mu\theta}=\varepsilon^{1/2}\stac{(1/2)}{F^{\mu
\theta}}+\cdots$. We will show that this ${\cal
O}(\varepsilon^{1/2})$ term in $F_{\mu\theta}$ also vanishes. The
bulk energy-momentum tensor contains terms like
$F_{\mu\lambda}F^{\nu\lambda}$ but these do not contribute to the
low energy effective theory as they are higher order in the
gradient expansion. The 5D Ricci tensor is given by
\begin{eqnarray}
{}^{5}\!R_{\mu}^{\;\nu} &=&\varepsilon \frac{1}{a^2}
\left(
R_{\mu}^{\;\nu}[h] -\cD_\mu \cD^\nu \tilde\psi
-\cD_\mu \tilde\psi
\cD^\nu \tilde\psi  \right)+\cdots,
\\
{}^{5}\!R_\theta^{\;\theta} &=&-\varepsilon \frac{1}{a^2}
\left( {\cal D}_\lambda {\cal D}^\lambda \tilde\psi
+ \cD_\lambda \tilde\psi
\cD^\lambda  \tilde\psi  \right)+ \cdots,
\end{eqnarray}
and ${}^5\! R_{\theta}^{\;\mu}={\cal O}(\varepsilon^{3/2})$, where
$\tilde\psi:=\psi^{(0)}-\lambda$. $R_{\mu}^{\;\nu}[h]$ and ${\cal
D}_\mu$ are respectively the Ricci tensor and the covariant
derivative constructed from $h_{\mu\nu}(x)$.

\subsection{Zeroth order equations}

The $\theta$ component of the Maxwell equations at zeroth order
reads
\begin{eqnarray}
\partial_y \Bigl(a^4 e^{-\phi^{(0)}+\psi^{(0)}}\stac{(0)}{F^{y \theta}}\Bigr)=0,
\end{eqnarray}
while the equation of motion for the dilaton at zeroth order is
given by
\begin{eqnarray}
\frac{1}{a^4}\py \Bigl(a^4f e^{\psi^{(0)}} \py \phi^{(0)} \Bigr)+\frac{1}{2}
\left(\frac{1}{M^2\ell}\stac{(0)}{F_{y\theta}}\right)^2e^{2\lambda-\phi^{(0)}-\psi^{(0) }}
-\frac{1}{2}e^{2\lambda+\phi^{(0)}+\psi^{(0)}}=0.
\end{eqnarray}

The $(\mu\nu)$ and $(\theta\theta)$ components of the evolution
equations are given respectively by
\begin{eqnarray}
&&f \left[ \py \left(\frac{\py a}{a} \right)
+\left(4 \frac{\py a}{a}+\frac{\py f}{f}+\py\psi^{(0)}  \right)
\frac{\py a}{a}\right]
\nonumber\\&&\qquad\qquad\qquad
=\frac{1}{4}\left(\frac{1}{M^2\ell}\stac{(0)}{F_{y\theta}}\right)^2e^{2\lambda-\phi^{(0)}-2\psi^{(0)}}
-\frac{1}{4}e^{2\lambda+\phi^{(0)}},
\\
&&f \left[\py \left(\frac{\py f}{2f}+\py \psi^{(0)} \right)
+ \left(4 \frac{\py a}{a}+\frac{\py f}{f}+\py \psi^{(0)}   \right)
\left(\frac{\py f}{2f}+\py \psi^{(0)}\right)    \right]
\nonumber\\&&\qquad\qquad\qquad
=-\frac{3}{4}\left(\frac{1}{M^2\ell}\stac{(0)}{F_{y\theta}}\right)^2e^{2\lambda-\phi^{(0)}-2\psi^{(0)}}
-\frac{1}{4}e^{2\lambda+\phi^{(0)}},
\end{eqnarray}
and
the Hamiltonian constraint becomes
\begin{eqnarray}
4f\left[3\left(\frac{\py a}{a}\right)^2+\frac{\py a}{a}\left(\frac{\py f}{f}+2\py\psi^{(0)}\right)\right]
= \left(\frac{1}{M^2\ell}\stac{(0)}{F_{y\theta}}\right)^2e^{2\lambda-\phi^{(0)}-2\psi^{(0)}}
+ f\bigl(\py \phi^{(0)}\bigr)^2-e^{2\lambda+\phi^{(0)}}.
\end{eqnarray}

The solutions for the above equations are obtained as
\begin{eqnarray}
a(y)=\sqrt{y},\qquad
f(y)=\frac{1}{4}\left(-y+\frac{\mu}{y}-\frac{q^2}{y^3} \right),
\qquad
\lambda(x) = \frac{1}{2}\Phi(x) ,
\end{eqnarray}
and
\begin{eqnarray}
\psi^{(0)}(y, x)=\Phi(x)+\sigma(x),
\qquad
\phi^{(0)}(y, x)=-\ln y-\Phi(x),
\qquad
\stac{(0)}{F}_{y\theta}= M^2\ell\frac{ q}{a^4}e^{\phi^{(0)}+\psi^{(0)}}
=M^2\ell \frac{q}{y^3}e^{\sigma(x)},
\end{eqnarray}
where $\mu$ and $q$ are integration constants. The momentum
constraint implies $\partial_{\mu}\sigma = 0$, and therefore
$\sigma=$ constant. This immediately leads to
$\stac{(1/2)}{F_{\mu\theta}}=0$ and hence $F_{\mu\theta}={\cal
O}(\varepsilon^{3/2})$. In the following, we put $\sigma=0$
without loss of generality. The 6D metric at the zeroth order is
given by
\begin{equation}
g_{MN} dx^M dx^N = e^{\Phi(x)} \left[ L_I^2 \frac{dy^2}{f} +
\ell^2 f d \theta^2 \right]  +a^2(y) h_{\mu \nu}(x) dx^{\mu}
dx^{\nu}.
\end{equation}
Then we can see that $\Phi(x)$ is associated with the scaling
symmetry $g_{MN} \to e^{\omega} g_{MN}$ and $e^{\phi} \to e^{\phi
- \omega}$. In fact, we will find that a solution for $h_{\mu
\nu}$ is given by $h_{\mu \nu} = e^{\Phi} \eta_{\mu \nu}$ if the
brane preserves the scaling symmetry, where $\eta_{\mu \nu}$
denotes the 4D Minkowski metric.

\subsection{First order equations}
At first order, the $(\mu\nu)$ component of the evolution equations is given by
\begin{eqnarray}
\frac{\sqrt{f}}{L_I}e^{-\Phi/2}\left[ \py
\!\stac{(1)}{K_{\mu}^{\;\nu}}+ \left(\frac{2}{y}+\frac{\py
f}{2f}\right)\!\stac{(1)}{K_{\mu}^{\;\nu}} +\frac{1}{2y}\Bigl(
\stac{(1)}{K_{\lambda}^{\;\lambda}}+\stac{(1)}{K_{\theta}^{\;\theta}}
\Bigr)\delta_{\mu}^{\;\nu} \right] &=&\frac{1}{y}\left(
R_{\mu}^{\;\nu}-{\cal D}_\mu {\cal D}^\nu \Phi-\frac{3}{2}{\cal
D}_\mu \Phi {\cal D}^\nu \Phi\right) \nonumber\\&&\qquad
-\frac{1}{4L_I^2}e^{\phi^{(0)}} \phi^{(1)} \delta_\mu^{\;\nu}
+\frac{1}{4} \cF\delta_{\mu}^{\;\nu}, \label{ev1st}
\end{eqnarray}
where
\begin{eqnarray}
\cF:=
\frac{1}{M^4}\Bigl(\stac{(0)}{F_{y\theta}}\stac{(1)}{F^{y\theta}}
+\stac{(1)}{F_{y\theta}}\stac{(0)}{F^{y\theta}}\Bigr)
e^{-\phi^{(0)}}-\frac{1}{M^4} \stac{(0)}{F}_{y \theta}
\stac{(0)}{F^{y\theta}}e^{-\phi^{(0)}}\phi^{(1)}.
\end{eqnarray}
The 4D Ricci tensor $R_{\mu}^{\;\nu}$ does not depend on $y$
because it is computed from $h_{\mu\nu}$ which is a function of
$x^{\mu}$ only and the index is raised by $h_{\mu \nu}$.

The 4D traceless part of Eq.~(\ref{ev1st}) is found to be
\begin{eqnarray}
\py\left(y^2\sqrt{f}\;\mathbb{K}_{\mu}^{\;\nu}\right)
= e^{\Phi/2}yL_I\mathbb{R}_{\mu}^{\;\nu},
\label{tl}
\end{eqnarray}
where
we defined
$\mathbb{K}_{\mu}^{\;\nu}:=\stac{(1)}{K_{\mu}^{\;\nu}}-(1/4)\delta_{\mu}^{\;\nu}\!\!\stac{(1)}{K_{\lambda}^{\;\lambda}}$
and
\begin{eqnarray}
\mathbb{R}_{\mu}^{\;\nu}:=R_{\mu}^{\;\nu}-\frac{1}{4}
\delta_{\mu}^{\;\nu}R-\left({\cal D}_\mu {\cal D}^\nu \Phi
-\frac{1}{4}\delta_{\mu}^{\;\nu} {\cal D}^2 \Phi \right)-\frac{3}{2}\left[
{\cal D}_\mu \Phi {\cal D}^\nu \Phi
-\frac{1}{4}\delta_{\mu}^{\;\nu} ({\cal D} \Phi)^2 \right],
\end{eqnarray}
where $\cD^2\Phi:=h^{\mu\nu}\cD_{\mu}\cD_{\nu}\Phi$
and $(\cD\Phi)^2:=h^{\mu\nu}\cD_{\mu}\Phi\cD_{\nu}\Phi$.
The general solution to the above equation is given by
\begin{eqnarray}
\mathbb{K}_{\mu}^{\;\nu}=\frac{e^{\Phi/2}}{2\sqrt{f}}L_I\mathbb{R}_{\mu}^{\;\nu}
+\frac{1}{y^2\sqrt{f}}\mathbb{C}_{\mu}^{\;\nu}(x),\label{general-sol}
\end{eqnarray}
where the traceless tensor $\mathbb{C}_{\mu}^{\;\nu}(x)$ is the integration ``constant''
to be fixed by the boundary conditions.

The 4D trace part of the evolution equations is
\begin{eqnarray}
\frac{\sqrt{f}}{L_I}e^{- \Phi/2} \left[ \py
\!\stac{(1)}{K_{\lambda}^{\;\lambda}}+\left(\frac{4}{y}+\frac{\py
f}{2f}\right)
\!\stac{(1)}{K_{\lambda}^{\;\lambda}}+\frac{2}{y}\stac{(1)}{K_{\theta}^{\;\theta}}
\right]=\frac{1}{y}\left[ R-{\cal D}^2 \Phi-\frac{3}{2}({\cal
D}\Phi)^2 \right] +
\cF-\frac{1}{L_I^2}\frac{e^{-\Phi}}{y}\phi^{(1)} , \label{tp1st}
\end{eqnarray}
and the $(\theta\theta)$ component of the evolution equations is
\begin{eqnarray}
\frac{\sqrt{f}}{L_I}e^{-\Phi/2}\left[ \py
\!\stac{(1)}{K_{\theta}^{\;\theta}}+ \left(\frac{2}{y}+\frac{\py
f}{f}\right)\!\stac{(1)}{K_{\theta}^{\;\theta}} +\frac{\py
f}{2f}\stac{(1)}{K_{\lambda}^{\;\lambda}} \right]
=-\frac{1}{2y}\left[{\cal D}^2 \Phi+({\cal D}\Phi)^2\right]
-\frac{3}{4} \cF-\frac{1}{4L_I^2 }\frac{e^{-\Phi}}{y}\phi^{(1)}.
\label{thth1st}
\end{eqnarray}
The Hamiltonian constraint at first order reduces to
\begin{eqnarray}
\frac{1}{y}\left[R-{\cal D}^2\Phi-\frac{3}{2}({\cal D}\Phi)^2
\right]+ \cF =2\frac{\sqrt{f}}{L_I}e^{-\Phi/2}\left[
\left(\frac{3}{2y}+\frac{\py
f}{2f}\right)\stac{(1)}{K_{\lambda}^{\;\lambda}}+
\frac{2}{y}\stac{(1)}{K_{\theta}^{\;\theta}}\right]+\frac{1}{L_I^2}\frac{e^{-\Phi}}{y}\phi^{(1)}
+ \frac{2f}{L_I^2}\frac{e^{-\Phi}}{y}\py\phi^{(1)}. \label{ham1st}
\end{eqnarray}
The dilaton equation of motion at first order reads
\begin{eqnarray}
&&\frac{f}{L_I^2}e^{-\Phi}\left[\py^2\phi^{(1)}+\left(\frac{2}{y}+\frac{\py
f}{f}\right)\py \phi^{(1)}\right]
-\frac{\sqrt{f}}{L_I}\frac{e^{-\Phi/2}}{y}\left(
\stac{(1)}{K_{\lambda}^{\;\lambda}}+\stac{(1)}{K_{\theta}^{\;\theta}}
\right) \nonumber\\&&\qquad\qquad\qquad\qquad\qquad\qquad
-\frac{1}{y}\left[\cD^2\Phi+(\cD\Phi)^2\right]
-\frac{1}{2L_I^2}\frac{e^{-\Phi}}{y}\phi^{(1)}+\frac{1}{2}\cF=0.\label{dil1st}
\end{eqnarray}

Now we define convenient quantities
\begin{eqnarray}
\cJ:=n^y\py\phi^{(1)}+\frac{1}{2}\stac{(1)}{K_{\lambda}^{\;\lambda}}
\end{eqnarray}
and
\begin{eqnarray}
\cK:=\frac{3}{4}
\stac{(1)}{K_{\lambda}^{\;\lambda}}+\stac{(1)}{K_{\theta}^{\;\theta}}
+\frac{\sqrt{f}}{L_I}e^{-\Phi/2}\left(\frac{\py
f}{2f}-\frac{1}{2y}\right)\psi^{(1)}+
\frac{y}{M^4\ell^2L_I\sqrt{f}}\stac{(0)}{F_{y\theta}}
e^{-\Phi/2}A_{\theta}^{(1)}
-\frac{\sqrt{f}}{L_I}\frac{e^{-\Phi/2}}{y}\phi^{(1)}.
\end{eqnarray}
The evolution equations for these variables can be derived
using Eqs.~(\ref{tp1st})--(\ref{dil1st}). With some manipulation one arrives at
\begin{eqnarray}
\py\left(y^2\sqrt{f}\cJ\right)&=&
\frac{1}{2}e^{\Phi/2}yL_I\left[
 R+\cD^2\Phi+\frac{1}{2}(\cD\Phi)^2
\right],\label{evJ}\\
\py\left(y^2\sqrt{f}\cK\right)&=&\frac{1}{4}e^{\Phi/2}yL_I\left[
 R-3\cD^2\Phi-\frac{7}{2}(\cD\Phi)^2
\right].\label{evK}
\end{eqnarray}
The two equations have the same structure as that of
Eq.~(\ref{tl}). The general solution for each evolution equation
contains one integration ``constant'' which will be determined by
the boundary conditions.

In terms of the above variables,
the momentum constraint equations are simplified to
\begin{eqnarray}
 \cD_{\nu}\left(e^{\Phi/2}\mathbb{K}_{\mu}^{\;\nu}\right)
- \cD_{\mu}\left(e^{\Phi/2}\cK\right)
+e^{\Phi/2}\cJ\cD_{\mu}\Phi=0.\label{momKJ}
\end{eqnarray}

\section{Junction conditions and effective theory on a regularized brane}\label{sec:branes}

\begin{figure}
\scalebox{.5}
 { \includegraphics*{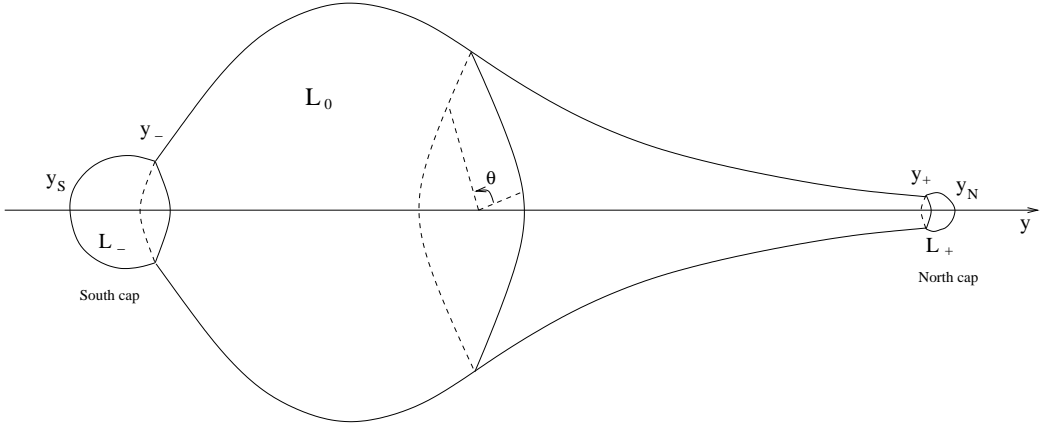}
  }
\caption{Schematic representation of the bulk spacetime with two
regularized caps.}\label{fig:fig}
\end{figure}

Our choice of parameters $\mu$, $q$ implies that $f(y)$ vanishes
at $y_N$ and $y_S$. These points are conical singularities that
are sourced by 3-branes. In order to accommodate usual matter on
the branes we need to resolve these singularities. We will use the
regularization scheme of~\cite{peloso1, uvcap}. The conical branes
are replaced with cylindrical codimension-one branes at $y=y_\pm$
and their interiors are filled with regular caps. See figure
\ref{fig:fig} for a sketch of the model. The geometry of the caps
and the central bulk is described by the 6D solutions found in the
previous section, with different curvature scales $L_+$ ($L_-$)
for the north (south) cap and $L_0$ for the central bulk.

The action of each brane is taken to be
\begin{eqnarray}
S_{{\rm brane}}=-\int d^5x\sqrt{-q}\left[
V(\phi)+\frac{1}{2}U(\phi)(\partial_{\hat\mu}\Sigma-\e A_{\hat\mu} )
(\partial^{\hat\mu}\Sigma-\e A^{\hat\mu} )\right]+\int d^5x\sqrt{-q}\,{\cal L}_{{\rm m}},
\end{eqnarray}
where $q_{\hat\mu\hat\nu}$ is the induced metric on the 4-brane,
$V(\phi)$ and $U(\phi)$ are the couplings to the dilaton,
and ${\cal L}_{{\rm m}}$ is the Lagrangian of usual matter localized on the brane.
At this stage we assume that the brane matter ${\cal L}_{{\rm m}}$
does not couple to the dilaton field.
We introduce a Stueckelberg field $\Sigma$, which is obtained by integrating out
the massive radial mode of a brane Higgs field.
The equation of motion for $\Sigma$ gives the gradient expansion form of the solution as~\cite{FKS}
\begin{eqnarray}
\Sigma(\theta, x)=n\theta+c^{(0)}(x)+\varepsilon c^{(1)}(x)+\cdots,
\end{eqnarray}
where $n$ must be an integer because of the periodicity $\theta\simeq\theta+2\pi$.

The jump conditions for the Maxwell field are
\begin{eqnarray}
\left[\left[ n^MF_{MN}e^{-\phi}\right]\right]=-\e U(\partial_N\Sigma-\e A_N),\label{gen-max}
\end{eqnarray}
while for the dilaton field we have
\begin{eqnarray}
\left[\left[ n^M\partial_M\phi \right]\right]=\frac{1}{M^4}\left[\frac{dV}{d\phi}
+\frac{1}{2}\frac{dU}{d\phi}
(\partial_{\hat \lambda}\Sigma -\e A_{\hat \lambda})(\partial^{\hat \lambda}\Sigma-\e A^{\hat \lambda})
\right],\label{gen-dil}
\end{eqnarray}
where $[[ F ]]_{y_{{\rm b}}}:=\lim_{\epsilon\to 0}\left(F|_{y_{{\rm b}}+\epsilon}-F|_{y_{{\rm b}}-\epsilon}\right)$.
Here and hereafter in this section all the quantities are evaluated at the position of the brane
under consideration.
The Israel conditions are given by
\begin{eqnarray}
\left[\left[ K_{\hat\mu}^{\;\hat\nu}-\delta_{\hat\mu}^{\;\hat\nu}\hat K\right]\right]
=-\frac{1}{M^4}T_{\hat \mu ({\rm tot})}^{\;\hat \nu}\label{gen-is}
\end{eqnarray}
where
\begin{eqnarray}
T_{\hat \mu ({\rm tot})}^{\;\hat \nu}=-V \delta_{\hat \mu}^{\;\hat \nu}
+U \left[
(\partial_{\hat \mu}\Sigma -\e A_{\hat \mu})(\partial^{\hat \nu}\Sigma-\e A^{\hat \nu})
-\frac{1}{2}\delta_{\hat \mu}^{\;\hat \nu}
(\partial_{\hat \lambda}\Sigma -\e A_{\hat \lambda})(\partial^{\hat \lambda}\Sigma-\e A^{\hat \lambda})
\right]+T_{\hat \mu}^{\;\hat \nu},
\end{eqnarray}
and $T_{\hat \mu}^{\;\hat \nu}$ represents the matter energy-momentum tensor.

\subsection{Zeroth order}

At zeroth order in the gradient expansion
the junction conditions~(\ref{gen-max})--(\ref{gen-is}) are written as
\begin{eqnarray}
\text{Maxwell:}&&\;\;
\left[\left[ \frac{\sqrt{f}}{L_I}\!\stac{(0)}{F}_{y \theta} y\,e^{\Phi/2}\right]\right]=-\e U^{(0)}
\left(n-\e A_\theta^{(0)}\right),\label{maxj1}
\\
\text{Dilaton:}&&\;\;
\left[\left[ \frac{\sqrt{f}}{L_I}\frac{1}{y}e^{-\Phi/2} \right]\right] =-\frac{1}{M^4}\left[
\frac{dV^{(0)}}{d\phi^{(0)}}+\frac{1}{2}\frac{dU^{(0)}}{d\phi^{(0)}}\frac{e^{-\Phi}}{\ell^2f}
\left(n-\e A_{\theta}^{(0)}\right)^2\right],\label{dilj1}
\\
\text{Israel\;}(\mu\nu):&&\;\;
\left[\left[ \frac{{\sqrt {f}}}{L_I} \left(\frac{3}{2y}+\frac{\py f}{2 f} \right)e^{-\Phi/2} \right]
\right]=-\frac{1}{M^4}\left[
V^{(0)} +\frac{1}{2}U^{(0)}  \frac{e^{-\Phi}}{\ell^2f}
\left(n-\e A_{\theta}^{(0)}\right)^2\right],\label{Imn1}
\\
\text{Israel\;}(\theta\theta):&&\;\;
\left[\left[ \frac{{\sqrt {f}}}{L_I} \frac{2}{y} e^{-\Phi/2}\right]
\right]=
-\frac{1}{M^4}\left[
V^{(0)} -\frac{1}{2}U^{(0)}  \frac{e^{-\Phi}}{\ell^2f}
\left(n-\e A_{\theta}^{(0)}\right)^2\right].\label{Ithth1}
\end{eqnarray}
The above conditions relate several
parameters with each other, and the detail of the parameter
counting of the configuration is found in Ref.~\cite{uvcap}. In
particular, the dilaton jump condition~(\ref{dilj1}) and the
Israel condition~(\ref{Ithth1}) imply
\begin{eqnarray}
\frac{V^{(0)}}{2}-\frac{dV^{(0)}}{d\phi^{(0)}}-\frac{1}{2}\frac{e^{-\Phi}}{\ell^2 f}
\left(\frac{U^{(0)}}{2}+\frac{dU^{(0)}}{d\phi^{(0)}}\right)
\left(n-\e A_{\theta}^{(0)}\right)^2=0.\label{lambda3}
\end{eqnarray}

The classical scaling symmetry is preserved by the special choice
of the potentials~\cite{sugra1, uvcap}
\begin{eqnarray}
V(\phi)=v e^{\phi/2}, \qquad U(\phi)=u e^{-\phi/2}.
\label{scalepotential}
\end{eqnarray}
With these potentials the junction
conditions~(\ref{maxj1})--(\ref{Ithth1}) put no constraints on
$\Phi(x)$ and Eq.~(\ref{lambda3}) is trivially satisfied. In this
case the first order analysis will provide the equation of motion
for $\Phi(x) $, as will be seen in the next subsection. In the
following, we assume that at the zeroth order, the potentials are
given by (\ref{scalepotential}), that is,
$U^{(0)}(\phi^{(0)})=u^{(0)}e^{-\phi^{(0)}/2}$ and
$V^{(0)}(\phi^{(0)})=v^{(0)}e^{\phi^{(0)}/2}$. Then we expand the
potentials as follows:
\begin{eqnarray}
V(\phi) &=& V^{(0)}(\phi^{(0)}) + \varepsilon
\left( V^{(1)}(\phi^{(0)}) + \frac{d V^{(0)}}{d \phi^{(0)}} \phi^{(1)} \right), \\
U(\phi) &=& U^{(0)}(\phi^{(0)}) + \varepsilon \left( U^{(1)}(\phi^{(0)}) + \frac{d
U^{(0)}}{d \phi^{(0)}} \phi^{(1)} \right),
\end{eqnarray}
where $V^{(1)}(\phi^{(0)})$ and $U^{(1)}(\phi^{(0)})$ stand for
the deviations from the zeroth order potentials.

\subsection{First order}

The 4D traceless part of the Israel conditions at first order is
given by
\begin{eqnarray}
\left[\left[\mathbb{K}_{\mu}^{\;\nu}\right]\right]=-\frac{1}{M^4}\mathbb{T}_{\mu}^{\;\nu},
\label{ktless}
\end{eqnarray}
where
$\mathbb{T}_{\mu}^{\;\nu}:=T_{\mu}^{\;\nu}-(1/4)\delta_{\mu}^{\;\nu}T_{\lambda}^{\;\lambda}$.
The 4D trace part of the Israel conditions reduces to
\begin{eqnarray}
\Biggl[\Biggl[\frac{3}{4}
\stac{(1)}{K_{\lambda}^{\;\lambda}}+\stac{(1)}{K_{\theta}^{\;\theta}}
\Biggr]\Biggr] = \frac{1}{4M^4}T_{\lambda}^{\;\lambda}
- \frac{1}{M^4} \Delta V +\frac{U^{(0)}}{M^4}\Delta
-\frac{1}{M^4}\left[ \frac{dV^{(0)}}{d\phi^{(0)}} +\frac{1}{2} \frac{dU^{(0)}}{d\phi^{(0)}}
\frac{e^{-\Phi}}{\ell^2f}
\left(n-\e A_{\theta}^{(0)}\right)^2
\right]\phi^{(1)},\label{tr1}
\end{eqnarray}
where we defined
\begin{equation}
\Delta V = V^{(1)}(\phi^{(0)}) + \frac{1}{2} U^{(1)}(\phi^{(0)}) \frac{e^{-\Phi}}{\ell^2 f}
\left (n-\e A_{\theta}^{(0)} \right)^2,
\end{equation}
and
\begin{eqnarray}
\Delta:=\frac{e^{-\Phi}}{\ell^2 f}\left(n-\e A_{\theta}^{(0)}\right)
\left[
\e A_{\theta}^{(1)}+
\left(n-\e A_{\theta}^{(0)}\right)\psi^{(1)}
\right].
\end{eqnarray}

Using the zeroth order junction conditions, Eq.~(\ref{tr1}) simply gives
\begin{eqnarray}
[[\cK]] =\frac{1}{4M^4}T_{\lambda}^{\;\lambda}-\frac{1}{M^4}
\Delta V.\label{ktr}
\end{eqnarray}
The $(\theta\theta)$ component of the Israel conditions is
\begin{eqnarray}
\Bigl[\Bigl[\stac{(1)}{K_{\lambda}^{\;\lambda}}\Bigr]\Bigr]&=&\frac{T_{\theta}^{\;\theta}}{M^4}
-\frac{U^{(0)}}{M^4}\Delta-\frac{1}{M^4}\left[
 \frac{dV^{(0)}}{d\phi^{(0)}} -\frac{1}{2} \frac{dU^{(0)}}{d\phi^{(0)}}
\frac{e^{-\Phi}}{\ell^2 f}\left(n-\e A_{\theta}^{(0)}\right)^2
\right]\phi^{(1)} \nonumber\\&&-\frac{1}{M^4}\Delta V+ U^{(1)}(\phi^{(0)})
\frac{e^{-\Phi}}{M^4\ell^2 f} \left (n-\e A_{\theta}^{(0)}
\right)^2,
\end{eqnarray}
and the dilaton jump condition is
\begin{eqnarray}
\left[\left[n^y\py\phi^{(1)}\right]\right] &=&
-\frac{1}{M^4}\frac{dU^{(0)}}{d\phi^{(0)}}\Delta+\frac{1}{M^4}\left[
 \frac{d^2V^{(0)}}{d\phi^{(0) 2}} +\frac{1}{2} \frac{d^2 U^{(0)}}{d \phi^{(0) 2}}
\frac{e^{-\Phi}}{\ell^2 f}\left(n-\e A_{\theta}^{(0)}\right)^2
\right]\phi^{(1)}\nonumber\\&&-\frac{1}{M^4}\frac{d}{d \Phi}
(\Delta V)-\frac{1}{2}U^{(1)}(\phi^{(0)}) \frac{e^{-\Phi}}{M^4\ell^2 f}
\left (n-\e A_{\theta}^{(0)} \right)^2.
\end{eqnarray}

Using the fact that the zeroth order potential have the scale
invariant forms (\ref{scalepotential}), the above two conditions
are combined to give
\begin{eqnarray}
\left[\left[\cJ\right]\right] = \frac{1}{2M^4}T_{\theta}^{\;\theta}
-\frac{1}{M^4} \frac{d}{d \Phi} (\Delta V) - \frac{1}{2 M^4}
\Delta V.
\end{eqnarray}
Therefore, the momentum constraints become
\begin{equation}
\cD_{\nu}\left(e^{\Phi/2}T_{\mu}^{\;\nu}
-e^{\Phi/2} \Delta V \delta^{\;\nu}_{\mu}\right)
=\left( \frac{1}{2} T_{\theta}^{\;\theta} -\frac{d}{d\Phi}
(\Delta V) - \frac{1}{2} \Delta V \right) e^{\Phi/2}\cD_{\mu}\Phi.
\end{equation}
In terms of the energy-momentum tensor integrated along the $\theta$-direction,
\begin{eqnarray}
\overline{T}_{\hat\mu}^{\;\hat\nu}:=2\pi\ell \sqrt{f}e^{\Phi/2}T_{\hat\mu}^{\;\hat\nu},
\end{eqnarray}
this can be rewritten as
\begin{eqnarray}
\cD_{\nu} \overline{T}_{\mu}^{\;\nu}
=\frac{1}{2}\overline{T}_{\theta}^{\;\theta}  \cD_{\mu}\Phi.
\end{eqnarray}

To fix the integration constants completely, we need the boundary
conditions at the north and south poles. Near a pole with the
coordinate $y=y_p$, where $p=\{N,S\}$, we have $f\sim y-y_p$. In
order for the evolution equations~(\ref{tl}), (\ref{evJ}), and
(\ref{evK}) to be regular at the poles, we require
\begin{eqnarray}
\mathbb{K}_{\mu}^{\;\nu},\;\cK,\;\cJ \lesssim |y-y_p|^{1/2}\to 0.
\end{eqnarray}
Now we can determine all the integration constants included in the
general solutions for $\mathbb{K}_{\mu}^{\;\nu}$, $\cK$ and $\cJ$.
Since the structure of the evolution equations and boundary
conditions are identical for these three variables, we summarize
the procedure to fix the integration constants in
Appendix~\ref{app:solve}, and here we focus on the resulting
effective theory on the brane.

Using Eqs.~(\ref{ktless}) and~(\ref{ktr})
together with the solution for $\mathbb{K}_{\mu}^{\;\nu}$ and $\cK$
in terms of $R$ and $\Phi$, we end up with the effective equations
\begin{eqnarray}
e^{\Phi}\left(R_{\mu}^{\;\nu}[q^+]-\frac{1}{2}\delta_{\mu}^{\;\nu}
R[q^+] - \Phi^{;\nu}_{;\mu}+
\delta_{\mu}^{\;\nu}\Phi^{;\lambda}_{;\lambda} - \frac{3}{2}
\Phi_{;\mu} \Phi^{;\nu}+\frac{5}{4} \delta_{\mu}^{\;\nu}
\Phi_{;\lambda} \Phi^{;\lambda} \right) =\kappa_+^2
\left(\overline{T}_{\mu}^{+\nu} - \overline{\Delta V}^+
\delta^{\nu}_{\mu} \right) +\frac{a_-^2 }{a_+^2}\kappa_-^2 \left(
\overline{T}_{\mu}^{-\nu} - \overline{\Delta
V}^-\delta^{\nu}_{\mu}
\right), \nonumber\\
\label{effective}
\end{eqnarray}
where the 4D gravitational couplings are defined as
\begin{eqnarray}
\kappa^2_{\pm}:=\frac{a_\pm^2}{2\pi\ell_*^2M^4},
\quad
\text{with}
\quad
\ell_*^2=\ell\int_{y_S}^{y_N}L_I ydy,
\end{eqnarray}
$;$ denotes a covariant derivative with respect to the induced
metric $q^+_{\mu \nu} =a_+^2 h_{\mu \nu}$, $R_{\mu \nu}[q^+]$ is
Ricci tensor computed from $q^+_{\mu \nu}$ and the potential
integrated along the $\theta$-direction is defined as
\begin{equation}
\overline{\Delta V} = 2 \pi \ell \sqrt{f} e^{\Phi/2} \Delta V.
\end{equation}

The first order equations for $\cJ$ give the equation of motion for $\Phi$:
\begin{eqnarray}
\left(e^{\Phi} \right)^{;\mu}_{;\mu} =\frac{\kappa^2_+}{4}\left(
\overline{T}_{\lambda}^{+\lambda}-\overline{T}_{\theta}^{+\theta}
+ 2 \frac{d}{d \Phi} (\overline{\Delta V}^+) - 4 \overline{\Delta V}^+
\right)
+\frac{a_-^2}{a_+^2}\frac{\kappa^2_-}{4}\left(
\overline{T}_{\lambda}^{-\lambda}-\overline{T}_{\theta}^{-\theta}+
 2 \frac{d}{d \Phi} (\overline{\Delta V}^-) - 4 \overline{\Delta V}^-
\right).
\end{eqnarray}

For simplicity let us ignore the matter energy-momentum tensor and the
potential on the south brane:
$\overline{T}^{-\hat\nu}_{\hat\mu}=\overline{\Delta V}^-=0$.
In the absence of the $(\theta\theta)$ component of the energy momentum tensor
on the north brane,
the 4D effective equations can be deduced from the action
\begin{eqnarray}
S_{{\rm eff}}=\int d^4x\sqrt{-q^+}\left[\frac{e^\Phi}{2
\kappa_+^2} \left(R[q^+]-\omega_{{\rm BD}} \Phi_{;\mu}\Phi^{;
\mu}\right)- \overline{\Delta V}^+ + \overline{{\cal L}}_{{\rm
m}}^+\right],
\end{eqnarray}
with the Brans-Dicke parameter $\omega_{{\rm BD}}=1/2$
(see also Appendix B of Ref.~\cite{KM2}).

\subsection{The exact time-dependent solutions in the 4D effective theory}\label{sec:4Det}
We now consider cosmological solutions in the 4D effective theory
and compare them with the known solutions to
the full 6D field equations~\cite{TBRH, KM2, CS}.

Let us assume $\overline{T}_{\mu}^{\pm \nu}=\overline{\Delta
V}^-=0$ and $\overline{T}_{\theta}^{\pm\theta}=0$. We consider the
case where the first order potential is scale invariant form. Then
$\Delta V^+ \propto e^{-\Phi/2}$ and $\overline{\Delta V}^+ =$
const. $:=\Lambda/\kappa_+^2$. We go to the Einstein frame defined
by
\begin{equation}
\tilde h_{\mu\nu}=e^{\Phi} q^+_{\mu\nu},
\end{equation}
and then the equations of motion become
\begin{eqnarray}
\tilde{R}_\mu^{\;\nu}[\tilde h]-\frac{1}{2}\delta_\mu^{\;\nu}\tilde{R}[\tilde h]
&=&-\Lambda
e^{-2\Phi}\delta_\mu^{\;\nu}+2
\Phi_{|\mu}\Phi^{|\nu}-\Phi_{|\lambda}\Phi^{|\lambda}\delta_\mu^{\;\nu},
\\
\Phi_{|\mu}^{\;|\mu}&=&-\Lambda e^{-2\Phi},
\end{eqnarray}
where $|$ stands for the covariant derivative with respect to the Einstein
frame metric $\tilde h_{\mu\nu}$.

Taking $\tilde h_{\mu\nu}$ to be a flat Friedman-Robertson-Walker
metric, $\tilde
h_{\mu\nu}dx^{\mu}dx^{\nu}=A^2(\tau)(-d\tau^2+d\mathbf{x}^2)$, the
equations of motion reduce to
\begin{equation}
\frac{A''}{A}=\frac{2}{3}\Lambda A^2e^{-2\Phi}-\frac{1}{3}\Phi'^2,
\quad \left(\frac{A'}{A}\right)^2=\frac{1}{3}\Lambda
A^2e^{-2\Phi}+\frac{1}{3}\Phi'^2,
\end{equation}
and
\begin{equation}
\Phi''+2\frac{A'}{A}\Phi'=\Lambda A^2e^{-2\Phi},
\end{equation}
where ${}':=d/d\tau$.
For $\Lambda=0$ a solution of the above equations is
\begin{equation}
A^2(\tau)=A_1\tau+A_2, \quad \Phi(\tau)=\pm\sqrt{3}\ln A(\tau) +
A_3,\label{power}
\end{equation}
where $A_i\;(i =1,2,3)$ are integration constants.
For $\Lambda\neq0$ a solution is
\begin{equation}
A(\tau)=e^{C \tau }, \quad \Phi(\tau)=\ln A(\tau),\label{exponential}
\end{equation}
where $C^2=\Lambda/2$. This indicates that $q^+_{\mu \nu} =
e^{\Phi} \eta_{\mu \nu}$ which is expected from the scaling
symmetry.

The brane scale factor $a_{{\rm b}}$ and the ``radion'' $\Psi$ are given by
\begin{equation}
a_{{\rm b}} = e^{-\Phi/2}A, \quad \Psi=e^{\Phi}.
\end{equation}
The solution (\ref{power}) gives the same 4D observables
(the brane scale factor and radion)
as an exact 6D solution found by Copeland and Seto
(equation~(70) of~\cite{CS}). In the same way the solution
(\ref{exponential}) reproduces the 4D quantities of their equation
(76). (The latter solution was first found by Tolley et al.~\cite{TBRH}.)
Thus we show that the solutions to the full 6D equations are reproduced by
our 4D effective theory on a regularized brane with the scale invariant potential
and with or without additional ``tension'' $\Lambda$.

At the zeroth order, the amplitudes $v^{(0)}$ and $u^{(0)}$ of the
potentials are fine-tuned. If we change the amplitude of the scale
invariant potentials, which is equivalent to add a cosmological
constant in the 4D effective theory, we get a runaway potential
for $\Phi$ and the 4D spacetime becomes non-static.

\subsection{Breaking the scale invariance}
Finally, let us consider the case where the first order potential
breaks the scale invariance. As the BD parameter is given by
$1/2$, this model violates the constraints coming from the solar
system experiments unless the BD scalar $\Phi$ is stabilized. It
is suggested that the potential on a brane can naturally stabilize
the modulus. For example, if we consider potentials
$V_1(\phi^{(0)}) = v^{(1)} e^{s \phi^{(0)}}$ and $U_1(\phi^{(0)})
= u^{(1)} e^{t \phi^{(0)}}$, the effective potential is given by
\begin{equation}
\overline{\Delta V} = 2 \pi \ell \sqrt{f} e^{\Phi/2} \left(
\frac{v^{(1)}}{y_+^s} e^{-s \Phi} + \frac{u^{(1)}}{2\ell^2fy_+^t}
e^{-(t+1) \Phi} (n- \e A^{(0)}_{\theta})^2 \right).
\end{equation}
As we saw in the previous subsection, if we take $s=1/2$ and
$t=-1/2$, $\overline{\Delta V}$ is independent of $\Phi$. However,
in general, it is possible to have a potential with a minimum by
choosing $s, t, v^{(1)}$ and $u^{(1)}$ appropriately \cite{uvcap}.
Then $\Phi$ can be stabilized and general relativity (GR) is
recovered.

\section{Conclusions}
In this paper, we derived the low energy effective theory in the
six-dimensional supergravity with resolved 4-branes. The gradient
expansion method is used to solve the bulk geometry. The resultant
effective theory is a Brans-Dicke theory with the Brans-Dicke
parameter given by $\omega_{\rm BD}=1/2$. If we choose the dilaton
potentials on the branes so that they keep the scaling symmetry in
the bulk and if we tune their amplitudes then there is no
potential in the effective theory and the modulus is massless.
Thus the static four-dimensional spacetime has vanishing
cosmological constant. It is also possible to obtain
time-dependent solutions due to the dynamics of the modulus field
and we showed that they are identified with the six-dimensional
exact time dependent solutions found in \cite{CS}.

Even if the potentials preserve the scaling symmetry, it was found
that there appears an effective cosmological constant in the
four-dimensional effective theory by changing the amplitude of the
potentials. Then in the Einstein frame, the modulus field acquires
an exponential potential and the static solution is no longer
allowed. Again, we showed that the cosmological solutions obtained
in the effective theory can be identified with the six-dimensional
exact time dependent solutions found in \cite{TBRH, KM2, CS}.

Our effective theory allows us to discuss cosmology with arbitrary
matter on the brane. As the BD parameter is given by $1/2$, it is
impossible to reproduce realistic cosmology without stabilizing
the modulus field. As it was suggested by Ref.~\cite{uvcap}, it is
easy to generate a potential for the modulus $\Phi$ with a minimum
by breaking the scaling symmetry from the dilaton potentials on
the branes. Then it is possible to reproduce GR at low energies.
However, once we stabilize the modulus, the cosmological constant
on a brane curves the four-dimensional spacetime in the same way
as in GR.

Our result would indicate that it is possible to reproduce
sensible cosmology in this six-dimensional supergravity model at
low energies but it would be difficult to address the cosmological
constant problem in this set-up. However, we should mention that
our effective theory is valid only up to the energy scale
determined by inverse of the size of extra-dimensions. This
condition is roughly given by $H \ell_* <1$ where $H$ is the
Hubble parameter. If we consider scales smaller than $\ell_*$
gravity becomes six-dimensional and from the table-top
experiments, $\ell_*$ is smaller than a few $\mu$m. Then for $H >
10^{-2}$ eV, our Universe becomes six-dimensional and it is
impossible to use the four-dimensional effective theory. In order
to address the behaviour of the universe at high energies, we
should deal with time-dependent solutions directly in
six-dimensional spacetime. This remains an open question.

Finally, we briefly make a comment on the limit where the 
codimension one branes are shrunk to codimension two objects. 
Our effective theory shows no pathological behaviour in this limit 
as long as the four-dimensional energy-momentum tensor integrated along 
the $\theta$-direction remains finite. However, 
in this limit, the first order extrinsic curvature 
$\!\stac{(1)}K_{\mu \nu}$ diverges and then the first order correction 
to the four-dimensional metric diverges. Then it is not clear whether there is 
a physical meaning in this limit. This is related to a deep issue 
of whether it is possible to put ordinary matter on codimension
2 objects \cite{matter} and we also leave this problem as an open 
question. 

\acknowledgments
We thank G. Tasinato for discussions. 
TK is supported by the JSPS under Contract No.~19-4199. TS is
supported by Grant-Aid for Scientific Research from Ministry of
Education, Science, Sports and Culture of Japan (Nos.~17740136,
17340075, and 19GS0219), the Japan-U.K., Japan-France and
Japan-India Research Cooperative Programs. KK is supported by
STFC. FA is supported by ``Funda\c{c}\~{a}o para a Ci\^{e}ncia e a
Tecnologia (Portugal)", with the fellowship's reference number:
SFRH/BD/18116/2004. The authors thank the Yukawa Institute for
Theoretical Physics at Kyoto University, where this work was
completed during the YITP-W-07-10 on "String phenomenology and
cosmology". 


\appendix

\section{Solving the bulk evolution equations}\label{app:solve}

All of the key evolution equations in the main text have the form of
\begin{eqnarray}
\py\left[y^2\sqrt{f} K(y, x) \right] = e^{\Phi(x)/2}y L_IR(x),
\end{eqnarray}
subject to the boundary conditions
\begin{eqnarray}
K(y_N, x)=K(y_S, x)=0,
\\
\left.[[K]]\right|_{y=y_{\pm}}=T^{\pm}(x).\label{app_junk}
\end{eqnarray}
In the south cap we have the solution
\begin{eqnarray}
K=\frac{y^2-y_S^2}{2y^2\sqrt{f}}e^{\Phi/2}L_-R.
\end{eqnarray}
In the bulk the solution can be written as
\begin{eqnarray}
K=\frac{e^{\Phi/2}\left[y^2L_0R+C(x)\right]}{2y^2\sqrt{f}},\label{Kb}
\end{eqnarray}
where the integration constant $C(x)$ is determined by the condition~(\ref{app_junk}) as
\begin{eqnarray}
C(x)= \left[-y_-^2L_0+(y_-^2-y_S^2)L_-\right]R+2y_-^2\sqrt{f_-}e^{-\Phi/2}T^-.\label{app:c=}
\end{eqnarray}
In the north cap we have the solution
\begin{eqnarray}
K=\frac{y^2-y_N^2}{2y^2\sqrt{f}}e^{\Phi/2}L_+R,
\end{eqnarray}
and the boundary condition~(\ref{app_junk}) requires
\begin{eqnarray}
\left[(y_N^2-y_+^2)L_++y_+^2L_0\right]R+C = -2y_+^2\sqrt{f_+}e^{-\Phi/2}T^+.\label{app:eq}
\end{eqnarray}
In the above we defined $f_{\pm}:=f(y_{\pm})$.
Substituting Eq.~(\ref{app:c=}) into Eq.~(\ref{app:eq}) we obtain
\begin{eqnarray}
\left(\int_{y_S}^{y_N}L_Iydy\right)\cdot R = -\sum_{i=\pm}y_i^2\sqrt{f_i}e^{-\Phi/2}T^i.
\end{eqnarray}

\section{${\cal O}(\varepsilon^{1/2})$ quantities}

The $\mu$ component of the ${\cal O}(\varepsilon^{1/2})$ Maxwell equations reads
\begin{eqnarray}
\py\left(
e^{-\phi^{(0)}}a^4 \stac{(1/2)}{F^{y\mu}}
\right)=0,
\end{eqnarray}
and thus we have
\begin{eqnarray}
\stac{(1/2)}{F^{ \mu y}} =M^2\frac{C_1^{\mu}(x)}{y^4}.
\end{eqnarray}
The ${\cal O}(\varepsilon^{1/2})$ evolution equation reduces to
\begin{eqnarray}
\frac{e^{-\Phi/2}}{L_I}\frac{1}{y^2}\py\left(y^2\sqrt{f}\stac{(1/2)}{K_{\theta}^{\;\nu}}\right)
&=&-\frac{1}{M^4}\stac{(0)}{F_{\theta y}}\stac{(1/2)}{F^{\nu y}}
\nonumber\\
&=&\ell q\frac{C_1^{\nu}}{y^6},
\end{eqnarray}
which can be integrated to give
\begin{eqnarray}
\stac{(1/2)}{K_{\theta}^{\;\nu}} = \frac{e^{\Phi/2}}{y^2\sqrt{f}}\left[
-\frac{L_I\ell q}{3}\frac{C_1^{\nu}(x)}{y^3}+C_2^{\nu}(x)\right].
\end{eqnarray}

The ${\cal O}(\varepsilon)$ evolution equations contain terms like
$\stac{(1/2)}{F_{\mu y}}\stac{(1/2)}{F^{\nu y}}\propto h_{\mu\lambda}C_1^{\lambda}C_1^{\nu}/f(y)$.
In the cap regions, we thus require $C_1^{\nu}=0$ to avoid the singular behavior at the poles.
Further, the regularity of $\stac{(1/2)}{K_{\theta}^{\;\nu}}$ at the poles
imposes $C_2^{\nu}=0$ in the cap regions.
To fix the integration constants in the central bulk, we use the
Maxwell jump conditions and Israel conditions at each brane:
\begin{eqnarray}
\left[\left[n_y\stac{(1/2)}{F^{y\mu }}e^{-\phi^{(0)}}\right]\right] &=&-\e U\left(
\partial^{\mu}\Sigma-\e A^{\mu}\right)^{(1/2)},
\\
\left[\left[\stac{(1/2)}{K_{\theta}^{\;\nu}}\right]\right]&=&-\frac{U}{M^4}\left(n-\e A_{\theta}^{(0)}\right)
\left(
\partial^{\mu}\Sigma-\e A^{\mu}\right)^{(1/2)}.
\end{eqnarray}
Combining these two equations and noting that $\stac{(1/2)}{F^{y\mu }}=0=\stac{(1/2)}{K_{\theta}^{\;\nu}}$
in each cap, we obtain two linear algebraic equations for
the bulk values of $C_1^{\nu}$ and $C_2^{\nu}$:
\begin{eqnarray}
\left.\left[\stac{(1/2)}{K_{\theta}^{\;\nu}}-\frac{1}{\e M^4}\left(n-\e A_{\theta}^{(0)}\right)
n_y\stac{(1/2)}{F^{y\mu }}e^{-\phi^{(0)}}\right]\right|_{y_{\pm}\mp\epsilon}=0.
\end{eqnarray}
Therefore, $C_1^{\nu}=C_2^{\nu}=0$ in the bulk.
Now we also see that
\begin{eqnarray}
\left(\partial^{\mu}\Sigma-\e A^{\mu}\right)^{(1/2)}=0 \quad \text{on the branes},
\end{eqnarray}
and $b_{\mu}={\cal O}(\varepsilon^{3/2})$.





\begin{thebibliography}{99}
\bibitem{sugra}
G.~W.~Gibbons, R.~Guven and C.~N.~Pope,
  Phys.\ Lett.\  B {\bf 595}, 498 (2004)
  [arXiv:hep-th/0307238];
  C.~P.~Burgess, F.~Quevedo, G.~Tasinato and I.~Zavala,
  JHEP {\bf 0411}, 069 (2004)
  [arXiv:hep-th/0408109];
  A.~J.~Tolley, C.~P.~Burgess, D.~Hoover and Y.~Aghababaie,
  JHEP {\bf 0603}, 091 (2006)
  [arXiv:hep-th/0512218].

\bibitem{sugra1}
Y.~Aghababaie {\it et al.},
  JHEP {\bf 0309}, 037 (2003)
  [arXiv:hep-th/0308064].

\bibitem{sugra2}
H.~M.~Lee and A.~Papazoglou,
  Nucl.\ Phys.\  B {\bf 747}, 294 (2006)
  [Erratum-ibid.\  B {\bf 765}, 200 (2007)]
  [arXiv:hep-th/0602208];
  C.~P.~Burgess, C.~de Rham, D.~Hoover, D.~Mason and A.~J.~Tolley,
  JCAP {\bf 0702}, 009 (2007)
  [arXiv:hep-th/0610078].

\bibitem{cc1}
J.~W.~Chen, M.~A.~Luty and E.~Ponton,
  JHEP {\bf 0009}, 012 (2000)
  [arXiv:hep-th/0003067];
   S.~M.~Carroll and M.~M.~Guica,
  arXiv:hep-th/0302067;
  I.~Navarro,
  JCAP {\bf 0309}, 004 (2003)
  [arXiv:hep-th/0302129];
 Y.~Aghababaie, C.~P.~Burgess, S.~L.~Parameswaran and F.~Quevedo,
  Nucl.\ Phys.\  B {\bf 680}, 389 (2004)
  [arXiv:hep-th/0304256].

\bibitem{cc-rev}
K.~Koyama,
  arXiv:0706.1557 [astro-ph].

\bibitem{burgess}
see C.~P.~Burgess,
  arXiv:0708.0911 [hep-ph]
for a recent review and references therein.

\bibitem{cc2}
I.~Navarro,
  Class.\ Quant.\ Grav.\  {\bf 20}, 3603 (2003)
  [arXiv:hep-th/0305014];
  H.~P.~Nilles, A.~Papazoglou and G.~Tasinato,
  Nucl.\ Phys.\  B {\bf 677}, 405 (2004)
  [arXiv:hep-th/0309042];
  H.~M.~Lee,
  Phys.\ Lett.\  B {\bf 587}, 117 (2004)
  [arXiv:hep-th/0309050];
  J.~Vinet and J.~M.~Cline,
  Phys.\ Rev.\  D {\bf 70}, 083514 (2004)
  [arXiv:hep-th/0406141];

  J.~Vinet and J.~M.~Cline,
  Phys.\ Rev.\  D {\bf 71}, 064011 (2005)
  [arXiv:hep-th/0501098].

\bibitem{GP}
J.~Garriga and M.~Porrati,
  JHEP {\bf 0408}, 028 (2004)
  [arXiv:hep-th/0406158].


\bibitem{Cline:2003ak}
  J.~M.~Cline, J.~Descheneau, M.~Giovannini and J.~Vinet,
  JHEP {\bf 0306}, 048 (2003)
  [arXiv:hep-th/0304147].

\bibitem{TBRH}
A.~J.~Tolley, C.~P.~Burgess, C.~de Rham and D.~Hoover,
  New J.\ Phys.\  {\bf 8}, 324 (2006)
  [arXiv:hep-th/0608083].

\bibitem{KM2}
  T.~Kobayashi and M.~Minamitsuji,
  arXiv:0705.3500 [hep-th].

\bibitem{CS}

  E.~J.~Copeland and O.~Seto,
  arXiv:0705.4169 [hep-th].


\bibitem{peloso1}
M.~Peloso, L.~Sorbo and G.~Tasinato,
  Phys.\ Rev.\  D {\bf 73}, 104025 (2006)
  [arXiv:hep-th/0603026].

\bibitem{ppz}
E.~Papantonopoulos, A.~Papazoglou and V.~Zamarias,
  JHEP {\bf 0703}, 002 (2007)
  [arXiv:hep-th/0611311].

\bibitem{peloso2}
B.~Himmetoglu and M.~Peloso,
  Nucl.\ Phys.\  B {\bf 773}, 84 (2007)
  [arXiv:hep-th/0612140].

\bibitem{KM}
  T.~Kobayashi and M.~Minamitsuji,
  Phys.\ Rev.\  D {\bf 75}, 104013 (2007)
  [arXiv:hep-th/0703029].

\bibitem{hybrid}
J.~Louko and D.~L.~Wiltshire,
  JHEP {\bf 0202}, 007 (2002)
  [arXiv:hep-th/0109099];
B.~M.~N.~Carter, A.~B.~Nielsen and D.~L.~Wiltshire,
  JHEP {\bf 0607}, 034 (2006)
  [arXiv:hep-th/0602086];
T.~Kobayashi and Y.~i.~Takamizu,
  arXiv:0707.0894 [hep-th];
S.~Kanno, D.~Langlois, M.~Sasaki and J.~Soda,
  arXiv:0707.4510 [hep-th];
  S.~A.~Appleby and R.~A.~Battye,
  arXiv:0707.4238 [hep-ph].

\bibitem{ppz_cos}
E.~Papantonopoulos, A.~Papazoglou and V.~Zamarias,
  arXiv:0707.1396 [hep-th].

\bibitem{ml_cos}
M.~Minamitsuji and D.~Langlois,
  arXiv:0707.1426 [hep-th].

\bibitem{FKS}
  S.~Fujii, T.~Kobayashi and T.~Shiromizu,
  arXiv:0708.2534 [hep-th].

\bibitem{E-M}
G.~W.~Gibbons and D.~L.~Wiltshire,
  Nucl.\ Phys.\  B {\bf 287}, 717 (1987)
  [arXiv:hep-th/0109093];
  S.~Mukohyama, Y.~Sendouda, H.~Yoshiguchi and S.~Kinoshita,
  JCAP {\bf 0507}, 013 (2005)
  [arXiv:hep-th/0506050].

\bibitem{soda}
S.~Kanno and J.~Soda,
  Phys.\ Rev.\  D {\bf 66}, 043526 (2002)
  [arXiv:hep-th/0205188].


\bibitem{sk}
T.~Shiromizu and K.~Koyama,
  Phys.\ Rev.\  D {\bf 67}, 084022 (2003)
  [arXiv:hep-th/0210066].

\bibitem{ge}
T.~Shiromizu, K.~Koyama, S.~Onda and T.~Torii,
  Phys.\ Rev.\  D {\bf 68}, 063506 (2003)
  [arXiv:hep-th/0305253];
  S.~Onda, T.~Shiromizu, K.~Koyama and S.~Hayakawa,
  Phys.\ Rev.\  D {\bf 69}, 123503 (2004)
  [arXiv:hep-th/0311262];
  F.~Arroja and K.~Koyama,
  Class.\ Quant.\ Grav.\  {\bf 23} (2006) 4249
  [arXiv:hep-th/0602068];
  K.~Koyama, K.~Koyama and F.~Arroja,
  Phys.\ Lett.\  B {\bf 641}, 81 (2006)
  [arXiv:hep-th/0607145];
  T.~Kobayashi, T.~Shiromizu and N.~Deruelle,
  Phys.\ Rev.\  D {\bf 74}, 104031 (2006)
  [arXiv:hep-th/0608166];
   K.~Koyama and K.~Koyama,
  Class.\ Quant.\ Grav.\  {\bf 22}, 3431 (2005)
  [arXiv:hep-th/0505256].

\bibitem{uvcap}
C.~P.~Burgess, D.~Hoover and G.~Tasinato,
  arXiv:0705.3212 [hep-th].

\bibitem{matter}
  N.~Kaloper and D.~Kiley,
  JHEP {\bf 0603}, 077 (2006)
  [arXiv:hep-th/0601110];
  C.~de Rham,
  arXiv:0707.0884 [hep-th].

\end{thebibliography}
\end{document}